\documentclass[%
 reprint,
superscriptaddress,
 amsmath,amssymb,
 aps,
]{revtex4-2}

\usepackage{graphicx}
\usepackage{dcolumn}
\usepackage{bm}
\usepackage{hyperref}
\hypersetup{
	colorlinks=true,
	linkcolor=blue,
	citecolor=magenta,
}

\begin{document}

\preprint{APS/123-QED}

\title{Anomalous normal state magnetotransport in an electron-doped cuprate}

\author{Nicholas R. Poniatowski}
\affiliation{Maryland Quantum Materials Center and Department of Physics, University of Maryland, College Park, Maryland 20742, USA}
\author{Tarapada Sarkar}
\affiliation{Maryland Quantum Materials Center and Department of Physics, University of Maryland, College Park, Maryland 20742, USA}
\author{Richard L. Greene}
\email{rickg@umd.edu}
\affiliation{Maryland Quantum Materials Center and Department of Physics, University of Maryland, College Park, Maryland 20742, USA}

\date{\today}

\begin{abstract}
We report magnetoresistance and Hall angle measurements of the electron-doped cuprate La$_{2-x}$Ce$_x$CuO$_4$ over a wide range of dopings from $x = 0.08 - 0.17$. Above 100 K, we find an unconventional $\sim H^{3/2}$ magnetic field dependence of the magnetoresistance observed in all samples doped within the superconducting dome. Further, the measured magnetoresistance violates Kohler's rule. Given the ubiquity of this anomalous magnetoresistance at high temperatures above the superconducting dome, we speculate that the origin of this behavior is linked to the unusual $\rho \sim T^2$ resistivity observed over the same wide parameter range at high temperatures. We also find a strong doping dependence of the Hall angle with an unconventional temperature dependence of $\cot \theta_H \sim T^{4}$ ($T^{2.5}$) for samples doped below (above) the Fermi surface reconstruction doping $x_{\text{FSR}} = 0.14$.

\end{abstract}

\maketitle

The normal state of the cuprate high-temperature superconductors has captivated the interest of the condensed matter physics community for the past decade while defying every attempt at theoretical explanation. The transport phenomena observed in these materials is believed to depart from the conventional Landau Fermi liquid theory of metals \cite{agd} and such ``non-Fermi liquid'' behaviors appear to be a common feature of disparate families of high-temperature superconductors \cite{Keimerreview}.  Consequently, it is reasonable to imagine that the mechanism of high-temperature superconductivity may naturally emerge from this non-Fermi liquid ``strange metal'' just as phonon-mediated superconductivity naturally emerges from a conventional Fermi liquid, making an understanding of this strange metallic state a potential stepping stone toward identifying the origin of high-temperature superconductivity. 

The typical hallmark of the strange metallic state is the infamous linear-in-$T$ resistivity of the hole-doped cuprates, which persists over an anomalously large temperature range, from $T_c$ to 1000 K in some systems \cite{martinbsco, lscotakagi}. However, the electron-doped compounds also exhibit a plethora of strange metallic behavior which differ sharply from the conventional properties of a Fermi liquid \cite{smreview}. Further, these materials display two distinct regimes of strange metallicity with different behaviors at high and low temperatures, and which may or may not be of a common origin.

The high-temperature strange metallic phase of the electron-doped cuprates is characterized by a universal quadratic-in-$T$ resistivity, seen in all compounds and for all dopings \cite{rickrmp, flstrangemetal} from roughly 100 K to above 600 K \cite{bach}. This $T^2$ behavior is what one might naively expect for a Fermi liquid. However, a more thoughtful consideration of the large magnitude of the resistivity and the high temperature scale at which it is observed lead one to conclude that is in fact an extremely strange transport behavior \cite{flstrangemetal}. In fact, it is arguably stranger than the high-temperature linear-in-$T$ resistivity of the hole-doped materials, which could potentially be explained by electron-phonon scattering in a low carrier density system \cite{sankarlineart} and is a generic feature of most conventional metals.

At low temperatures, the nature of the strange metallic ground state is strongly doping-dependent and non-universal. Generically, the phase diagram of the electron-doped cuprates is dominated by a Fermi surface reconstruction (FSR) which occurs inside the superconducting dome \cite{fsr0, fsr1, fsr2, fsr3, He3449} and is believed to be driven by short-ranged antiferromagnetic order \cite{fsrth} or the onset of topological order \cite{Sachdev_2018}. For the material of interest in this study, La$_{2-x}$Ce$_x$CuO$_4$ (LCCO), this FSR occurs at cerium concentration $x = .14$ \cite{taraprb} (for reference, the SC dome extends from $x = .07$ to $x = .175$). In samples doped below the FSR (i.e. $x < .14$ for LCCO), the low-temperature resistivity exhibits an upturn, increasing with decreasing temperature \cite{taraprb, pccoupturn}. The origin of this upturn, also seen in hole-doped cuprates \cite{originalupturn, upturn2}, is not well understood, but is thought to be associated with the underdoped materials' proximity to an antiferromagnetic insulating phase \cite{upturnspin,lscoupturn}.

The low temperature transport behavior of samples doped beyond the FSR ($x > .14$ in LCCO) has proven to be particularly intriguing. Remarkably, the resistivity in this region of the phase diagram varies linearly with temperature from a doping-dependent crossover temperature of the order of tens of Kelvin down to the lowest measured temperature of 30 mK when superconductivity is suppressed with an external magnetic field \cite{kuinature}. This is in stark contrast to the Fermi liquid expectation of a low-temperature $T^2$ resistivity, and is perhaps the most compelling evidence available for a non-Fermi-liquid ground state. In addition, it has recently been found that the low-temperature magnetoresistance (MR) of these overdoped samples is linear in magnetic field, in contrast to the conventional $H^2$ dependence expected for weak fields \cite{tarasciadv}. 

\begin{figure*}
    \centering
    \includegraphics[width=180 mm]{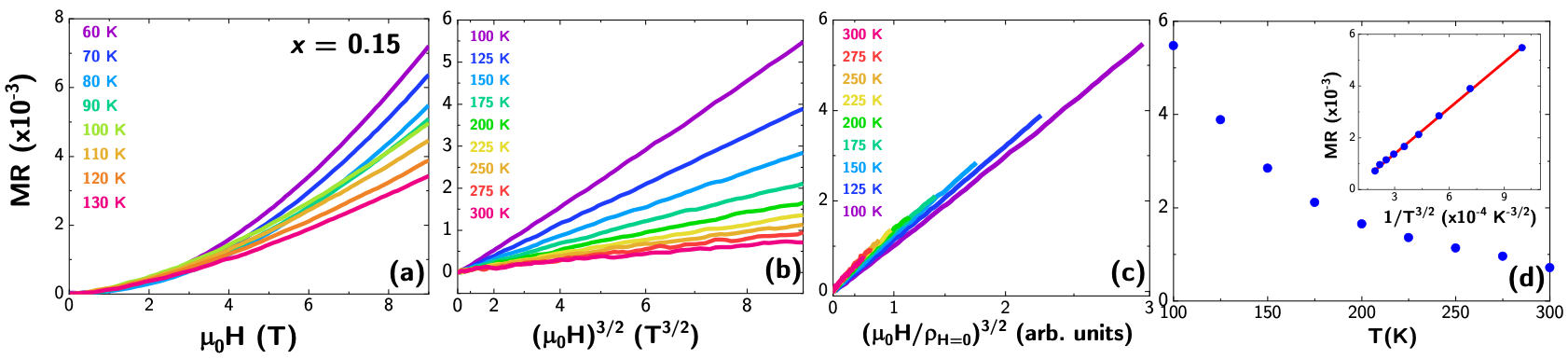}
    \caption{Magnetoresistance (MR) data for an $x = 0.15$ LCCO sample measured up to 9 T. (a) raw MR $= [\rho(T,H) - \rho(T,0)]/\rho(T,0)$ curves for temperatures between 60 - 130 K. The MR is quadratic in field below 80 K \cite{tarasciadv}, but the crossing of the curves taken at 80, 90, and 100 K suggests the MR at 80 K grows more rapidly with field than at 100 K. That is, MR $\sim H^n$ with $n < 2$ at 100 K and above. (b) The MR from 100 - 300 K plotted against $(\mu_0 H)^{3/2}$. The linearity of the curves clearly implies that MR $\sim H^{3/2}$. (c) Kohler's scaling plot of the MR from (b) plotted against $[\mu_0 H/\rho(T,0)]^{3/2}$. Collapse of all the data onto one curve would suggest Kohler's rule is obeyed. (d) The magnitude of the MR at 9 T plotted as a function of temperature; inset: The same data plotted against $1/T^{3/2}$. The linearity of the data on this scale suggests that MR$(T) \sim 1/T^{3/2}$.}
    \label{x15}
\end{figure*}

In light of this correspondence between the temperature-dependent resistivity and MR in the overdoped strange metallic ground state, in this work we investigate whether a similar correspondence exists for the high-temperature metallic state where $\rho \sim T^2$. To this end, we report measurements of the low-field transverse MR ($H \perp ab$-plane) for LCCO samples spanning the phase diagram, from $x = 0.08$ to $x = 0.17$. 
All measurements are performed on $c$-axis oriented epitaxial thin films of LCCO grown via pulsed laser deposition on SrTiO$_3$ substrates. Details of the sample preparation can be found in the literature \cite{taraprb}. 

In Fig. \ref{x15}, we show representative MR $= [\rho(T,H) - \rho(T,0)]/\rho(T,0)$ data for an $x = 0.15$ LCCO sample. In Fig. \ref{x15}a, one sees that the curves taken at 80, 90 and 100 K cross one another, indicating the MR increases more rapidly with field at 80 K than at 100 K. Further, by inspection the MR at lower temperatures appears to be quadratic in field, while the higher temperature curves seem to have a weaker field dependence. This is confirmed in Fig. \ref{x15}b, where we plot the MR from 100 K to room temperature as a function of $(\mu_0 H)^{3/2}$. The linearity of the curves on this scale clearly shows that the MR scales as MR $\sim H^{3/2}$, in contrast to the typical weak-field behavior of MR $\sim H^2$ anticipated for conventional metals.

\begin{figure}
    \centering
    \includegraphics[width=90mm]{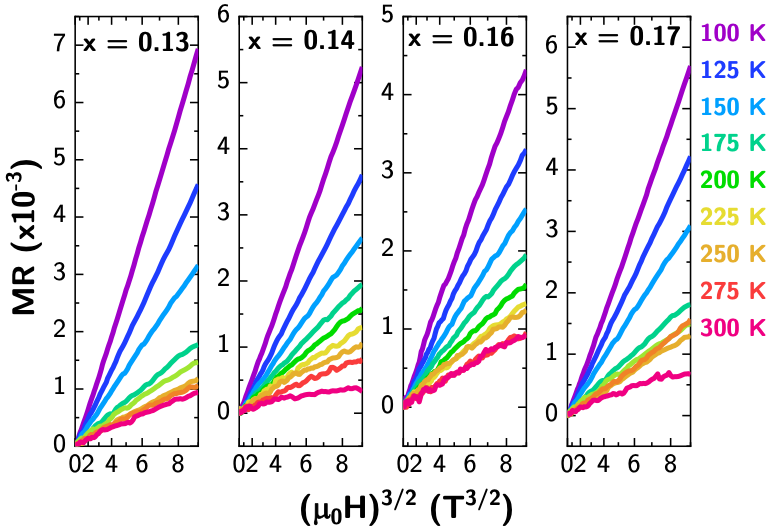}
    \caption{Plots of the magnetoresistance (MR) vs. $(\mu_0 H)^{3/2}$ for dopings $x = 0.13, 0.14, 0.16$, and $0.17$ at numerous temperatures between 100 - 300 K. The MR $\sim H^{3/2}$ behavior is observed for each doping. }
    \label{hscaling}
\end{figure} 

A useful lens through which to consider the MR behavior of a metallic system is Kohler's rule, the statement that the MR should depend on the ratio of the mean free path to the cyclotron radius in a simple semiclassical picture. More formally, as can be seen from the Boltzmann equation, it is the statement that the MR depends only on the product of the magnetic field and scattering time, or more practically (since $\rho(T,0) \sim \tau^{-1}$) is a function of only the ratio of the magnetic field to the zero-field resistivity \cite{Ziman}, i.e. MR $= F[H/\rho(T,0)]$ for some function $F(x)$. In Fig. \ref{x15}c, we attempt to assess the validity of Kohler's rule in this system by plotting the MR against $(\mu_0 H)^{3/2}$, where the near-collapse of the data onto a single curve appears to suggest that Kohler's rule is obeyed. 

However, since MR $\sim H^{3/2}$ and $\rho(T,0) \sim T^2$, a corollary to Kohler's rule is that the temperature dependence of the MR (at fixed field) must be MR $\sim H^{3/2} \sim [H/\rho(T,0)]^{3/2} \sim 1/T^3$. The temperature-dependent MR at 9 T is plotted in Fig. \ref{x15}d, which, as can be seen in the inset, follows a clear MR $\sim 1/T^{3/2}$ behavior, in violation with Kohler scaling. That is, despite what visually appears to be a reasonable collapse of the data in Fig. \ref{x15}c, the temperature dependence of the MR is incompatible with Kohler's rule. The failure of Kohler's rule in LCCO is hardly surprising, given the extreme simplicity of the theory and the complexity of the material under study.

\begin{figure}
    \centering
    \includegraphics[width=90mm]{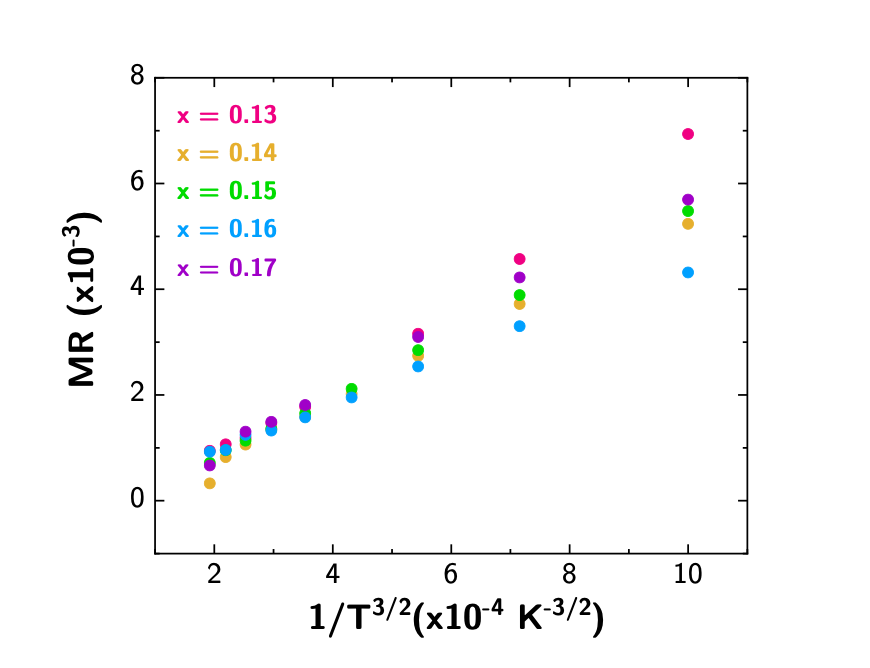}
    \caption{Temperature-dependent magnetoresistance (MR) at 9 T for dopings $x = 0.13-0.17$ plotted against $1/T^{3/2}$.}
    \label{tscaling}
\end{figure}

\begin{figure}
    \centering
    \includegraphics[width=90mm]{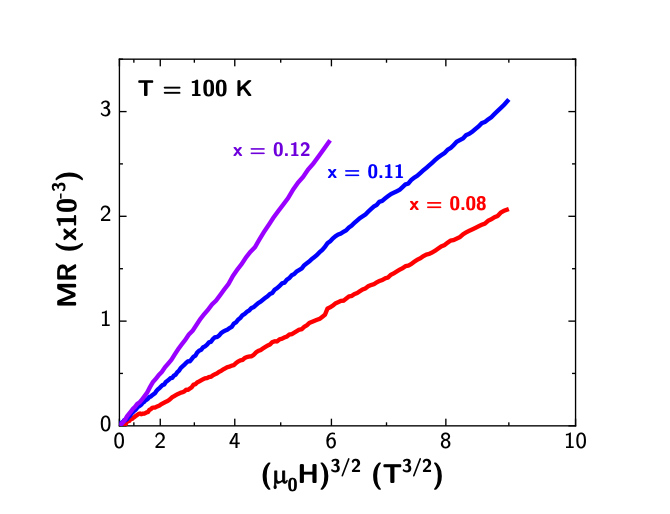}
    \caption{Magnetoresistance (MR) at 100 K plotted against $(\mu_0 H)^{3/2}$ for several dopings below the Fermi surface reconstruction doping $x_{\text{FSR}} = 0.14$. Again, the MR $\sim H^{3/2}$ behavior is seen for all dopings.}
    \label{lowdopings}
\end{figure}

Moving on from this single doping ($x = 0.15$), we plot the MR against $(\mu_0 H)^{3/2}$ for dopings $0.13 < x < 0.17$ in  Fig. \ref{hscaling}, and find that the MR $\sim H^{3/2}$ behavior is generic above 100 K. Further, in Fig. \ref{tscaling}, we show the temperature dependence of the MR is also $\sim 1/T^{3/2}$ for all dopings studied. Consequently, we find that Kohler's rule is violated across the LCCO phase diagram. 
 
\begin{figure}
    \centering
    \includegraphics[width=90mm]{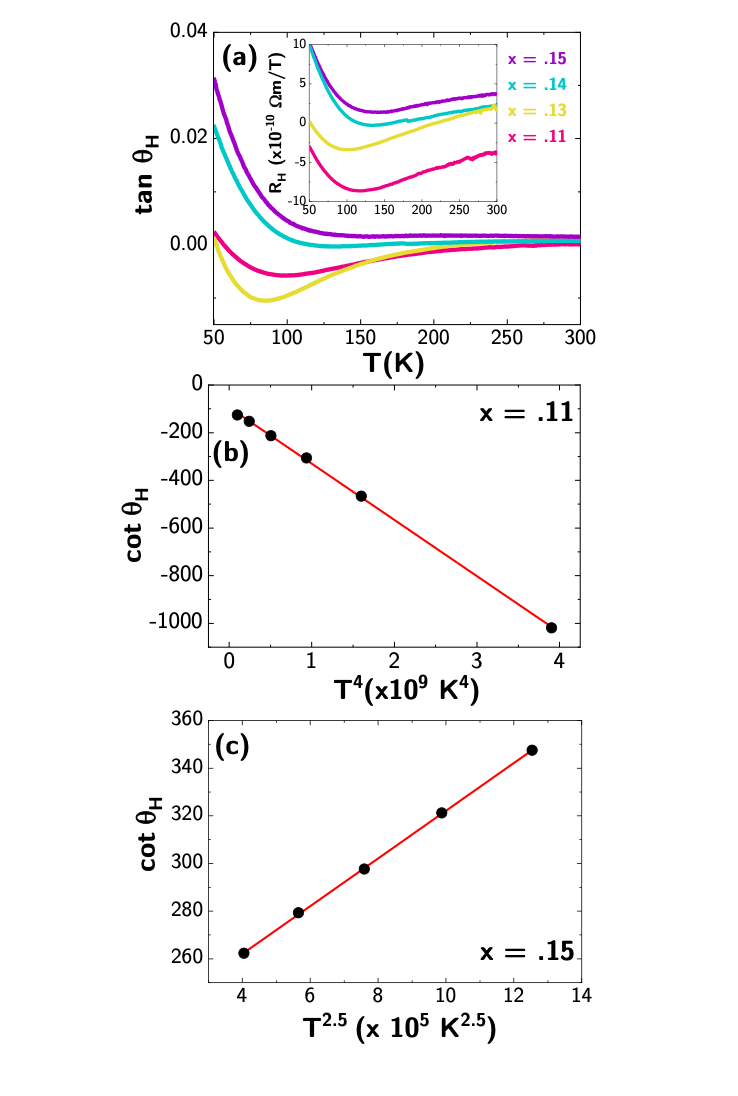}
    \caption{High-temperature Hall response: (a) $\tan \theta_H = \rho_{xy}/\rho_{xx}$ vs. temperature for several dopings of LCCO from 50 to 300 K; inset: Hall number for the same samples from 50 to 300 K. (b) $\cot \theta_H = \rho_{xx}/\rho_{xy}$ for the $x = 0.11$ sample from 150 to 300 K, plotted against $T^4$. The red line is a linear fit. (c) $\cot \theta_H$ for the $x = 0.15$ sample from 150 K to 300 K, plotted against $T^{2.5}$. The red line is a linear fit. }
    \label{hallfig}
\end{figure} 
 
To further illustrate the ubiquity of the MR $\sim H^{3/2}$ scaling at high-temperatures, we plot the MR at 100 K for several dopings well below the FSR doping $x_{\text{FSR}} = 0.14$, where the MR $\sim H^{3/2}$ dependence is clearly seen. The temperature-dependent MR was not studied in detail for these dopings.

Taken together, Figs. \ref{x15} - \ref{lowdopings} establish a universal $\sim H^{3/2}$ magnetic field dependence of the MR for $T \gtrsim $ 100 K across the LCCO phase diagram  from $x = 0.08 - 0.17$, i.e. for all dopings with a superconducting ground state. That is, the anomalous MR power law occurs over the same wide region of the phase diagram where the anomalous high-temperature $\rho \sim T^2$ dependence of the zero-field resistivity is observed. It is then natural to speculate that these two unconventional transport phenomena are driven by a common resistive scattering mechanism.

To supplement our MR measurements, in the inset of Fig. \ref{hallfig}a we present measurements of the of the Hall coefficient for several dopings throughout the superconducting dome. Note that even at room temperature, the Hall coefficient has a nontrivial temperature dependence, departing from single-carrier Fermi liquid expectations. Moreover, $R_H$ changes sign at high temperatures for dopings near the FSR, which naively suggests that both electron- and hole-like carriers may be relevant to the high-temperature transport properties of LCCO within this doping range. However, given the longstanding confusion over the meaning of the Hall coefficient in the cuprates, such a conclusion may very well be premature.

To further our characterization of the high-temperature transport phenomenology, in Fig. \ref{hallfig}b the tangent of the Hall angle, $\tan \theta_H \equiv \rho_{xy}/\rho_{xx}$ is shown for several dopings up to room temperature. Although the cotangent of the Hall angle, $\cot \theta_H = \rho_{xx}/\rho_{xy}$ is the typical quantity of theoretical interest, the zeroes of $\rho_{xy}$ which, as mentioned above, are present in some dopings up to 100-200 K, prevent the evaluation of this ratio for every doping. 

For the $x = 0.11$  and $x = 0.15$ samples, $\rho_{xy}$ does not change sign above 100 K, allowing for $\cot \theta_H$ to be analyzed. As shown in Figs. \ref{hallfig}b and \ref{hallfig}c,  $\cot \theta_H$ exhibits a $T^4$ temperature dependence for the $x = 0.11$ sample and a $T^{2.5}$ temperature dependence for the $x = 0.15$ sample. In contrast, a Fermi liquid is expected to have $\rho_{xx} \sim T^2$ and $\rho_{xy} \sim T^0$, and thus $\cot \theta_H \sim T^2$, which is in fact one of the defining features of a Fermi liquid, and a behavior observed in the hole-doped cuprates \cite{ongcothall, andocothall}. Moreover, the strong doping dependence of $\cot \theta_H$ (that is, the very different power laws exhibited on either side of the FSR) contrasts sharply with the largely doping independent longitudinal response. 

Altogether, we have demonstrated an anomalous $\sim H^{3/2}$ field-dependence of the MR in LCCO which coexists with the well-known $\rho \sim T^2$ temperature-dependent resistivity at high temperatures for all dopings within the superconducting dome. We hope that this new aspect of the high-temperature metallic state of the electron-doped cuprates will help identify the resistive scattering mechanism responsible for these unconventional transport phenomena, and perhaps its relationship to high-temperature superconductivity and/or the low-temperature strange metallic state. In contrast, we find the transverse response to be strongly doping dependent, with drastically different temperature dependences of the Hall angle for samples doped above or below the FSR doping $x_{\text{FSR}} = 0.14$.

\begin{acknowledgments}
We thank Sankar Das Sarma for helpful conversations regarding our results. This work is supported by the NSF under Grants No. DMR-1708334 and DMR-2002658, and the Maryland Quantum Materials Center. 
\end{acknowledgments}

\providecommand{\noopsort}[1]{}\providecommand{\singleletter}[1]{#1}%

\end{document}